\def\tit#1{``#1,''}
\long\def\comment#1{}

\let\ox=\otimes

\newcommand{\Mx}[3]{\left#1\begin{array}{rrrr}#2\end{array}\right#3}

\newcommand{\ket}[1]{| #1 \rangle}
\newcommand{\bra}[1]{\langle #1 |}

\newcommand{\vxv}[1]{\ket{#1}\bra{#1}}
\newcommand{\brkt}[2]{\langle #1 {\mid} #2 \rangle}

\newcommand\Hil{\mathcal H}
\newcommand\Ctrl{\mathsf {C_{trl}}}
\newcommand\Shft{\mathsf {S_{hft}}}

\newcommand{\GE}{\geqslant}

\newcommand{\op}[1]{\bm{#1}}
\newcommand{\wave}[1]{\hbox to #1{\leaders\hbox{${\sim}\!$}\hfil}}
\newcommand{\contr}[2]{\ensuremath{\mathsf C{\searrow}^{\!\!\!\!#1}_{#2}}}

\newcommand\eq[1]{Eq.~(\ref{#1})}
\newcommand\Sec[1]{Sec.~\ref{sec:#1}}
\documentclass[twocolumn]{article}
\usepackage{exscale}
\usepackage{amssymb,amsbsy,bm}
\begin{document}
\makeatletter
\renewcommand\section{\@startsection {section}{1}{\z@}%
                                   {-3.5ex \@plus -1ex \@minus -.2ex}%
                                   {2.3ex \@plus.2ex}%
                                   {\normalfont\large\bfseries}}
\renewcommand\subsection{\@startsection{subsection}{2}{\z@}%
                                     {-3.25ex\@plus -1ex \@minus -.2ex}%
                                     {\normalfont\bfseries}}
\makeatother
\textheight 240mm
\textwidth  165mm
\columnsep  5mm
\parindent 0pt
\parskip 1em
\oddsidemargin -4mm
\evensidemargin -4mm
\topmargin -15mm

\title{Quantum Processors and Controllers}
\author{{\em Alexander Yu.\ Vlasov}
\thanks{Federal Radiological Center (IRH),
197101 Mira Street 8, St.--Petersburg, Russia}
\thanks{
A. Friedmann Lab. Theor. Physics,
191023, Griboedov Can.\ 30/32, St.--Petersburg, Russia
}}
\date{Jan, May 2003}
\sloppy
\maketitle
\begin{abstract}
\noindent
In this paper is presented an abstract theory of quantum processors and
controllers, special kind of quantum computational network defined
on a composite quantum system with two parts: the controlling and
controlled subsystems. Such approach formally differs from
consideration of quantum control as some external influence on a system
using some set of Hamiltonians or quantum gates. The model of programmed
quantum controllers discussed in present paper is based on theory of
universal deterministic quantum processors (programmable gate arrays).
Such quantum devices may simulate arbitrary evolution of quantum system
and so demonstrate an example of universal quantum control.

\end{abstract}
\section{Introduction}
\label{sec:intro}
Let us consider simple example of control using a Hilbert space of composite
quantum system with two parts:
\begin{equation}
 \Hil = \Hil_c \ox \Hil_d.
\label{H12}
\end{equation}
Here $\Hil_d$ is a Hilbert space of quantum system considered as {\em data},
(controlled variables, subject of control) and $\Hil_c$ is Hilbert space
of {\em control} (``manager'', program of changes). The approach is close
analogue of {\em conditional quantum dynamics} \cite{Joz95}.

It is possible to start with {\em classical} example with
{\em reversible} {\sf Controlled-NOT} gate defined on set of two binary
variables as $(a,b) \to (a, a~{\sf XOR}~b)$, i.e., if first binary variable
is $a={\sf 0}$, then second one is unchanged, but if $a={\sf 1}$, then
$b \to {\sf NOT}~b$.
Reversible logical gates are usual tools in
quantum computations; here {\sf XOR} is
{\em eXclusive OR}, $a + b\ (\mathrm{mod}~2)$.
Construction of quantum {\sf Controlled-NOT} gate
is straightforward (see \cite{Cle99} or any other introduction in quantum
information science).

In quantum networks bits are changed to qubits (quantum
bits) and yet another quantum gate with two qubits is {\em controlled-U}
gate \cite{Cle99}, when quantum gate $U$ is applied to second qubit if
first one is $\ket{1}$ ($\ket{a}$ is notion for state of qubit in Dirac
notation), but if first qubit is $\ket{0}$ then second one is unchanged.

It is possible to write {\em controlled-U} as $4 \times 4$ matrix
\begin{equation}
\Mx({1&0&0&0\\0&1&0&0\\0&0&U_{11}&U_{12}\\0&0&U_{21}&U_{22}}),
\label{CtrlU}
\end{equation}
where $U_{ij}$ are components of quantum gate for second qubit, i.e.,
$2 \times 2$ matrix $U$.

To describe development of the idea, {\em programmable quantum gate arrays}
are used in \Sec{qga}. Such quantum devices also are called {\em quantum
processors}, but may be used as {\em quantum controllers} as well, it is
discussed in \Sec{qc}. More formal mathematical description of
{\em programmable quantum controllers} provided in \Sec{formal}.
Universality of quantum computations and control are briefly recollected
in \Sec{univer}. Some discussion on universal control with continuous
quantum variables is presented in \Sec{infin}.

\section{Programmable quantum gate arrays}
\label{sec:qga}
It is possible to use decomposition \eq{H12} with more general quantum
networks and describe process of control as
\begin{equation}
\Ctrl \colon \bigl(\ket{C} \ox \ket{\Psi}\bigr)
 \mapsto \ket{C'} \ox (\op u_C\ket{\Psi}),
\label{qctrl}
\end{equation}
i.e., $\Ctrl$ is some {\em fixed} network and different control strategies
ensured by different states $\ket{C}$ of control registers: for each such
state different operator $\op u_C$ acts on second system. The expression
\eq{qctrl} coincides with definition of special kind of quantum network,
{\em programmable quantum gate array} \cite{NC97,VMC01}.
Let us use notation \contr mn for $\dim\Hil_c=m$ and $\dim\Hil_d=n$.

It should be mentioned, that $\Ctrl$ as any other quantum computational
network \cite{Deu89} with pure states must be described as {\em linear
unitary operator} acting on the Hilbert space \eq{H12} (similarly with example
\eq{CtrlU} above for simplest case with two qubits). Here the quantum laws
have serious implications denoted already in \cite{NC97}: any two states
of ``program'' (first, control register) must be orthogonal, i.e., maximal
number of different operators $\op u_C$ availiable for application to second,
controlled system is equal to dimension of Hilbert space $\Hil_c$, i.e.,
for universal control $\dim\Hil_c = \infty$, because number of different
quantum gates is infinite.

To explain this result, let us consider two different ``control strategies''
$\ket{A}$ and $\ket{B}$
\begin{eqnarray*}
\Ctrl \bigl(\ket{A} \ox \ket{\Psi}\bigr)
 &=& \ket{A'} \ox (\op u_A\ket{\Psi}), \\
\Ctrl \bigl(\ket{B} \ox \ket{\Psi}\bigr)
 &=& \ket{B'} \ox (\op u_B\ket{\Psi}),
\end{eqnarray*}
but because $\Ctrl$ is unitary operator, it may not change scalar product
of two vectors, i.e.,
\begin{equation}
 \brkt AB =
 \brkt {A'}{B'}\, \bra{\Psi} \op u_A^\dag \op u_B \ket{\Psi}
\label{scprog}
\end{equation}

In \eq{scprog} $\brkt AB$ and $\brkt {A'}{B'}$ are fixed numbers, but for
$\op u_A \neq \op u_B$ term
$\bra{\Psi} \op u_A^\dag \op u_B \ket{\Psi}$ depends on $\ket\Psi$. But
\eq{scprog} must be satisfied for any $\ket\Psi$ and so
\begin{equation}
\brkt AB = \brkt {A'}{B'} = 0,
\end{equation}
i.e., states corresponding to different programs are {\em orthogonal}.

For example even for one controlled qubit, set of all possible gates may be
described by continuous three-dimensional family, i.e., even for this simple
case with $\dim\Hil_d = 2$, for universal control it is necessary to have
$\dim\Hil_c~=~\infty$ with control register described by three continuous
quantum variables (\contr{\infty^3}2).

It is interesting, that such enormous difference between size of control
and controlled system is rather subtle property of quantum dynamics,
for example, it may be found {\em linear}, but {\em non-unitary} operator
like \eq{qctrl} for universal control and with size of control register only
in two times bigger than for controlled quantum system \contr{n^2}n, it is
simply operator of multiplication of a matrix on a vector written as formal
linear map
\{$(\Hil \otimes\Hil^*) \otimes \Hil \to (\Hil \otimes\Hil^*) \otimes \Hil$;
$\op A \otimes v \mapsto \op 1 \otimes \op A v$)\},
but it would contradict to laws of quantum mechanics.

\section{Quantum processors as controllers}
\label{sec:qc}

It is clear, that such approach has some difference with other methods
of consideration of quantum control \cite{LV01,SGRR02,ZL03}, there controlled
system is also described as some state, but control is introduced as set of
``external'' controlling operators; gates or Hamiltonians,
i.e., control and data are described from different points of view
({\em semiclassical coherent control}).

\begin{itemize}
\item[{\em Note}]
\small
Another attempt of joint quantum description of
control and controlled system, using same term ``quantum controller,''
was included in \cite{Lloyd97}, as some perspective for
above-mentioned semiclassical coherent control. It was not suggested
a general model of such joint quantum description, but few illustrative
examples were presented.
But here is discussed an alternative approach, it is enough to recall
some distinctive principles of consistent framework for quantum control
with pure states considered in present paper:
\begin{enumerate}
 \item Control and controlled system {\em must not be entangled}.
  It follows directly from definition \eq{qctrl}.
 \item The consequence of such definition is {\em impossibility to use
  superposition of states in control register}.
 \item So, there is {\em noticeable asymmetry between
  control and controlled system in such approach.}
 \item Final development of the principles is {\em original three-level
  design of programmable quantum controller} discussed below in
  \Sec{formal}, Fig.~\ref{Fig3Buses}.
\end{enumerate}
\end{itemize}

The construction of programmable quantum gate arrays, or {\em quantum
processors} \cite{NC97,VMC01,Vla01a,Vla01b,HBZ01,HBZ02,Vla02} provides more
unified description of control and controlled system. It is in agreement
with principles just mentioned above. Term
``quantum controller'' may be also justified for such a system, because for
universal quantum processor on controlled system $\Hil_d$ formally may be
simulated practically any {\em physical process}, if to use tradition of
consideration of such systems suggested by R.~Feynman and D.~Deutsch
\cite{FeySim,Deu85}. For classical processors difference in sizes of program
and data is not such a radical and this new property of quantum processor
(controller) is related with infinite amount of different quantum programs
(algorithms of control).

Despite of discussed above result about infinite-dimensional controlling
register, universal quantum controllers with finite dimension of control
space $\Hil_c$ are also quite usefull. It should be mentioned
first so-called {\em stochastic quantum processors}
\cite{NC97,VMC01,HBZ01,HBZ02}. Such quantum processor does not produce
correct answer each time, but provides special ``check bit'' displaying
if answer is correct or not. If answer is not correct, it is suggested to
perform calculations again and again. Probability of correct answer is
reduced with size of data register and increases with number of tries.

Seems idea of stochastic quantum processor quite interesting, but has lot
of problems, for example it is not even clear if it is possible to use
composition of such networks for few-steps process due to unspecified
time of each step and it is certainly some problem for application of
such system as quantum controllers. It is also not quite clear, if it
is always possible to ``discard'' incorrect result of action for general
controller and start all again.

In addition, the ``ideal limit'' of
such design resembles non-possible linear (but non-unitary) operator
discussed earlier and it is similar with some other known models of
quantum systems (``relaxation'' gate, ``instantaneous'' reduction, etc.),
then balance between ``arduous'' and ``impossible'' is too fine and linked
with deep problems of quantum mechanics.

Anyway, the idea of stochastic gates seems useful, for example in
\cite{Vla02} was shown, that continuous limit of some special stochastic
network discussed in \cite{VMC01,HBZ02} coincides with continuous limit of
some ``deterministic'' quantum gate, despite of very different behavior in
finite, discrete case. It should be mentioned also, that main efforts of
many authors last time were applied rather to the stochastic design,
but deterministic one seems more appropriate for present consideration of
quantum controllers.

Another construction with finite control register uses ``universality
in approximate sense''. It is quite reasonable approach and based on idea,
that in realistic tasks always possible instead of continuous infinite
space of parameters to use only finite set of points for approximation.
The more dense set, the more accurate such a method. Some basic papers
about universality in quantum computation uses such approach
\cite{Deu85,Ek95,Gate95}.

\section{More rigor mathematical\\ description}
\label{sec:formal}

Let us consider quantum processors and controllers with more details
\cite{Vla01a,Vla01b}. It was already mentioned, that all different
states of controlling register must be orthogonal. Let us use for simplicity
finite controlling register and choose such orthogonal states as new basis.
It is possible to denote it simply as $\ket{0}$, $\ket{1}$, $\ket{2}$,
\dots, i.e., ``no operation,'' ``operation \#1,'' ``operation \#2,'' \dots

If $\dim \Hil_c = m$ and $\dim \Hil_d = n$ in \eq{H12}, then $\dim \Hil = mn$
and in suggested new basis $\Ctrl$ may be written as block-diagonal
$mn \times mn$ matrix
\begin{equation}
\Ctrl = \Mx({\op u_0\\&\op u_1&&\smash{\mbox{\Huge$0$}}\\
                 &&\ddots\\\smash{\mbox{\Huge$0$}}&&&\op u_{m-1}}),
\label{CtrlMat}
\end{equation}
where $\op u_k$ are $n \times n$ matrices and it is convenient to choose
$\op u_0 = \op 1$ (``no operation''). It is example of conditional
quantum dynamics described in \cite{Joz95} and using Dirac notation
it may be rewritten as \cite{Joz95}
\begin{equation}
\Ctrl = \vxv{0}\ox \op u_0 + \vxv{1}\ox \op u_1 + \cdots
\label{CtrlDir}
\end{equation}

\begin{figure*}[t]
\begin{center}
\unitlength=1mm
\begin{picture}(140.00,85)
\thinlines
\put(5,65.00){\dashbox{2}(50,20)[cc]{Pseudo-classical bus ($\Hil_p$)}}
\put(5,35.00){\dashbox{2.00}(50,20.00)[cc]{Intermediate bus ($\Hil_c$)}}
\put(5.00,5.00){\dashbox{2}(50.00,20.00)[cc]{Quantum bus ($\Hil_d$)}}
\thicklines
\put(55.00,10.00){\line(1,0){45}}
\put(55.00,11.00){\line(1,0){45}}
\put(55.00,12.00){\line(1,0){45}}
\put(125.00,10.00){\line(1,0){15}}
\put(125.00,11.00){\line(1,0){15}}
\put(125.00,12.00){\line(1,0){15}}
\put(55.00,18.00){\line(1,0){45.00}}
\put(55.00,19.00){\line(1,0){45.00}}
\put(55.00,20.00){\line(1,0){45.00}}
\put(125.00,18.00){\line(1,0){15.00}}
\put(125.00,19.00){\line(1,0){15.00}}
\put(125.00,20.00){\line(1,0){15.00}}
\put(55,40.00){\line(1,0){10.00}}
\put(55,41.00){\line(1,0){10.00}}
\put(55,42.00){\line(1,0){10.00}}
\put(90,40.00){\line(1,0){10}}
\put(90,41.00){\line(1,0){10}}
\put(90,42.00){\line(1,0){10}}
\put(125.00,40.00){\line(1,0){15}}
\put(125.00,41.00){\line(1,0){15}}
\put(125.00,42.00){\line(1,0){15}}
\put(55,48.00){\line(1,0){10}}
\put(55,49.00){\line(1,0){10}}
\put(55,50.00){\line(1,0){10}}
\put(90.00,48.00){\line(1,0){10.00}}
\put(90.00,49.00){\line(1,0){10.00}}
\put(90.00,50.00){\line(1,0){10.00}}
\put(125.00,48.00){\line(1,0){15.00}}
\put(125.00,49.00){\line(1,0){15.00}}
\put(125.00,50.00){\line(1,0){15.00}}
\put(55.00,70.00){\line(1,0){10}}
\put(55.00,71.00){\line(1,0){10}}
\put(55.00,72.00){\line(1,0){10}}
\put(90.00,70.00){\line(1,0){50}}
\put(90.00,71.00){\line(1,0){50}}
\put(90.00,72.00){\line(1,0){50}}
\put(55.00,78.00){\line(1,0){10}}
\put(55.00,79.00){\line(1,0){10}}
\put(55.00,80.00){\line(1,0){10}}
\put(90,78.00){\line(1,0){50}}
\put(90,79.00){\line(1,0){50}}
\put(90,80.00){\line(1,0){50}}
\put(65,35.00){\framebox(25,50.00)[cc]{
 \shortstack{Reversible\\program\\($\Shft$)}}}
\put(100,5.00){\framebox(25,50.00)[cc]{
 \shortstack{Quantum\\controller\\($\Ctrl$)}}}
\put(60,75.00){\makebox(0,0)[cc]{$\cdots$}}
\put(60,45.00){\makebox(0,0)[cc]{$\cdots$}}
\put(77.50,15.00){\makebox(0,0)[cc]{$\cdots$}}
\put(112.50,75.00){\makebox(0,0)[cc]{$\cdots$}}
\put(132.50,45.00){\makebox(0,0)[cc]{$\cdots$}}
\put(132.50,15.00){\makebox(0,0)[cc]{$\cdots$}}
\end{picture}
\end{center}

\caption{Design of programmable quantum controller with three buses
 ({\em cf} \cite{Vla01b}).}
\label{Fig3Buses}
\end{figure*}
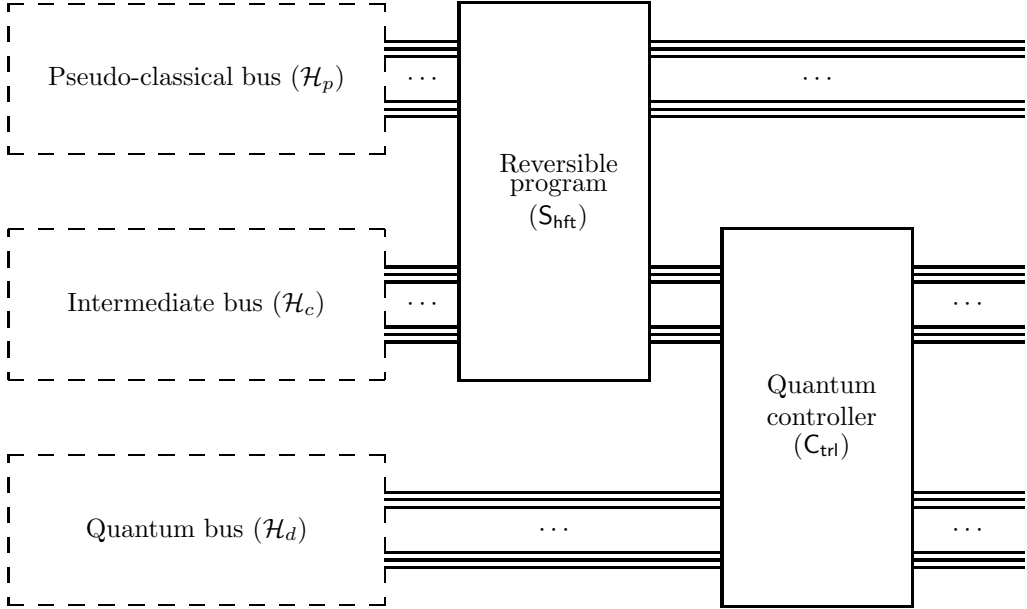

Such approach may be appropriate for simple quantum controller, but for
more difficult operations it is reasonable to consider an advanced design
\cite{Vla01a,Vla01b} of quantum processor that can be used
as {\em a programmable quantum controller}.
Instead of two systems \eq{H12} here is used design with three ``buses''
\begin{equation}
 \Hil = \Hil_p \ox \Hil_c \ox \Hil_d.
\label{H123}
\end{equation}
Here $\Hil_p,\Hil_c,\Hil_d$ are Hilbert spaces of {\em program, controller
and data}, or {\em pseudo-classical, intermediate and quantum buses}
respectively (see Fig~\ref{Fig3Buses}).

The idea is to use composition of two operators. First one was described
earlier, it is quantum controller $\Ctrl$ acting on intermediate and
quantum buses $\Hil_c \ox \Hil_d$ and second one acts on $\Hil_p \ox \Hil_c$
and on each step provides intermediate bus $\Hil_c$ with new state
$\ket{k}$ used as index $k$ of operator $\op u_k$ by quantum controller.

Let us consider simplest example with ``cyclic memory (ROM)''. Let
$\dim \Hil_c=m$ and it is necessary to perform program with $p$ steps.
Then $\dim \Hil_p=m^{p-1}$ and element $\ket{K}$ of $m^p$-dimensional
Hilbert space $\Hil_p \ox \Hil_c$ may be described as
\begin{equation}
 \ket{K} \equiv \ket{k_p,\ldots,k_2;k_1}
\end{equation}
and ``program'' is simply operator of cyclic shift
\begin{equation}
 \Shft\colon\ket{K}\mapsto\ket{k_1,k_p,\ldots,k_3;k_2}.
\end{equation}
Finally, for $p$ steps of {\em the programmable quantum controller}
with cyclic ROM (\contr{m^p}n), it is possible to write
\begin{equation}
 (\Shft\Ctrl)^p \colon \bigl(\ket{K}\ket{\Psi}\bigr) \mapsto
 \ket{K}(\op u_{k_p}\cdots\op u_{k_1}\ket{\Psi})
\label{CtrlShft}
\end{equation}
and because set of operators $\op u_k$ contains identity (unit), it is
possible to implement any sequence with up to $p$ operators using different
programs $\ket{K}$.

One problem here is huge size of program register. A method to
reduce the size is to use instead of shifted array more complicated
algorithm for generation of indexes. For example, instead of each sequence of
$n$ equivalent indexes $k$ it could use pair $(n,k)$.
It should be mentioned yet, that only reversible algorithms are appropriate
here due to common principles of quantum computations --- otherwise dynamics
would not be unitary. Really there are some methods of automatic conversion
of any algorithm to reversible one, but in such a case each step generates a
``garbage'' and size of program register may be even bigger, than for ROM. So
the area is related with classical theory of optimal reversible computations.

On the other hand, it was already mentioned earlier, that all states of
program register are orthogonal. It is not necessary to use superposition of
different states. It was a reason to call the register ``pseudo-classical.''
Such systems may be more simple for implementation \cite{Tsif} and so problem
with size may be not such essential, as for data register.

Yet another advantage of such pseudo-classical program register is
possibility to use ``halt bit'' and algorithms with variable length.
It is mentioned here, because such an opportunity is not very common for
general quantum algorithms due to quantum parallelism and interference
of different branches.

\section{Universality of quantum control}
\label{sec:univer}

When method of generation of arbitrary sequence of operators like
\eq{CtrlShft} is given, ideas of implementation of universal control
follows to standard procedures \cite{Cle99,Deu89,LV01,Vla01a,Vla01b,%
Deu85,Ek95,Gate95,DiVin95,UnSim}.

Let us consider case with finite size of control register. For good
approximation it is possible to choose $\op u_k$ near identity
operator, i.e.
\begin{equation}
\op u_k(\varepsilon) = \exp i \op H_k \varepsilon \approx
1 + i \op H_k \varepsilon \quad (\varepsilon \to 0).
\end{equation}
Here $\op H_k$ are Hermitian operator and corresponds to Hamiltonians
in some other approach to quantum control \cite{LV01}. Then small parameter
$\varepsilon$ is analogue of minimal time of action of some external
influence by the control Hamiltonian.

Due to general theory it is enough to have posibility to generate full
Lie algebra of Hermitian operators as linear span of $\op H_k$ and all
possible consequent commutators, but this part coinsides with general theory
of universal quantum computations and control and does not discussed
here in details
\cite{Cle99,Deu89,LV01,Vla01a,Vla01b,Deu85,Ek95,Gate95,DiVin95,UnSim,%
VlaUCl,VlaUNt}.

Algorithms of generation of indexes for application of different gates
$\op u_k$ often may be described using few nested cycles with repeating
series of states-indexes $\ket{k}$ \cite{Vla01a,VlaUCl,VlaUNt} due to general
algebraic approach with Lie algebras and commutators mentioned above.

Despite of such analogue in mathematical expressions, discussed approach
has some advantages due to closed description of controlling and controlled
systems. Really the quantum controller uses only one fixed Hamiltonian
$\op H_C$, $\Ctrl = e^{i \op H_C\,\delta t}$ and $\op H_k$ are rather formal
operators.

On the other hand, such quantum description has some difficulties, because
despite of pseudo-classical character of program register, it is anyway some
special kind of quantum system. It is not look reasonable, that for control
of such program register may be used some standard silicon chip. It was already
mentioned, that for presented model only reversible programs are
compatible with laws of quantum mechanics. Really it could be
simply shown, that irreversible operator for some state of quantum
controller has absurd property: it ``shrinks'' to zero wave vector of
the system (and all environment, due to linearity). One method of more
correct modeling is to use {\em mixed states} and density matrices.
But it is not only problem of given approach, most other models of
interface between classical devices and quantum system always provides
some challenge and may be much more nontrivial \cite{hybr2}.

\section{Continuous quantum variables}
\label{sec:infin}

Quantum computations with continuous variables is also active area of research
\cite{LlBr98}. The ideas presented here also possible to use in case of
control described by continuous quantum variables. For such a case direct
sum in \eq{CtrlDir} should be changed to direct integral \cite{Vla02}.
For such a system quantum control variables are continuous, but controlled
system is described by finite-dimensional Hilbert space. It is particular
case of {\em hybrid quantum computing} \cite{Llo00}.

Here pseudo-classical character of {\em program bus} provides some
simplification. It may be described using classical terms and it is in
agreement with relative success of usual semiclassical description of
quantum control. But {\em intermediate and quantum buses} may not be
considered using only classical ideas. Here {\em intermediate bus} could
provide some challenge as an ``interface'' between classical and
quantum world. In presented approach it is not so critical, because
{\em pseudo-classical bus} is also described as a quantum system and
was called so due to ``recommendation'' to use here only orthogonal set of
states. It is principally possible to apply any superposition of states to
such ``pseudo-classical'' bus, but in such a case states of control and
controlled system became {\em entangled} after application of
$\Ctrl$ \eq{qctrl} and it is not considered as
prescribed functioning of considered device.

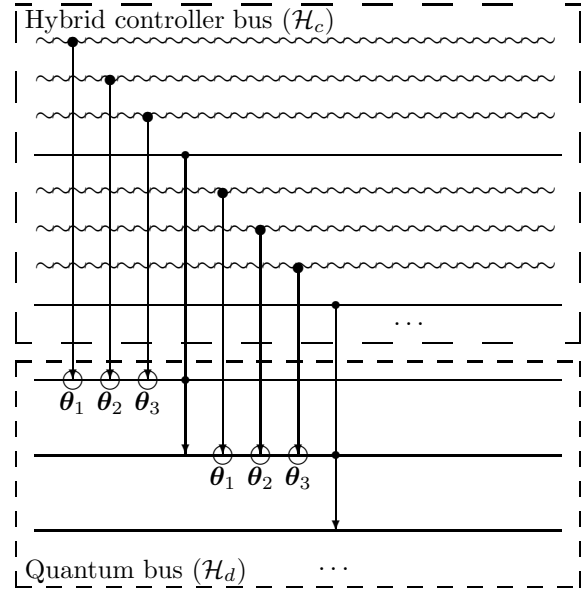
\begin{figure}[ht]
\begin{center}
\unitlength=1mm
\linethickness{0.4pt}
\begin{picture}(85,80)(5,0)
\put(10,75){\makebox(0,0)[lc]{\wave{7cm}}}
\put(10,70){\makebox(0,0)[lc]{\wave{7cm}}}
\put(10,65){\makebox(0,0)[lc]{\wave{7cm}}}
\put(10,60){\line(1,0){70}}
\put(10,55){\makebox(0,0)[lc]{\wave{7cm}}}
\put(10,50){\makebox(0,0)[lc]{\wave{7cm}}}
\put(10,45){\makebox(0,0)[lc]{\wave{7cm}}}
\put(10,40){\line(1,0){70}}
\put(10,30){\line(1,0){70}}
\put(10,20){\line(1,0){70}}
\put(10,10){\line(1,0){70}}
\put(15,75){\vector(0,-1){45}}
\put(15,75){\circle*{1.5}}
\put(20,70){\vector(0,-1){40}}
\put(20,70){\circle*{1.5}}
\put(25,65){\circle*{1.5}}
\put(25,65){\vector(0,-1){35}}
\put(30,60){\vector(0,-1){40}}
\put(30,60){\circle*{1}}
\put(30,30){\circle*{1}}
\put(35,55){\vector(0,-1){35}}
\put(35,55){\circle*{1.5}}
\put(40,50){\vector(0,-1){30}}
\put(40,50){\circle*{1.5}}
\put(45,45){\vector(0,-1){25}}
\put(45,45){\circle*{1.5}}
\put(50,40){\vector(0,-1){30}}
\put(50,40){\circle*{1}}
\put(50,20){\circle*{1}}
\put(60,37.5){\makebox(0,0)[cc]{\ldots}}
\put(50,5){\makebox(0,0)[cc]{\ldots}}
\put(15,30){\circle{2.5}}
\put(15,27){\makebox(0,0)[cc]{$\op\theta_1$}}
\put(20,30){\circle{2.5}}
\put(20,27){\makebox(0,0)[cc]{$\op\theta_2$}}
\put(25,30){\circle{2.5}}
\put(25,27){\makebox(0,0)[cc]{$\op\theta_3$}}
\put(35,20){\circle{2.5}}
\put(35,17){\makebox(0,0)[cc]{$\op\theta_1$}}
\put(40,20){\circle{2.5}}
\put(40,17){\makebox(0,0)[cc]{$\op\theta_2$}}
\put(45,20){\circle{2.5}}
\put(45,17){\makebox(0,0)[cc]{$\op\theta_3$}}
\put(7.5,35){\dashbox{4}(75,45)[lt]{ Hybrid controller bus
 $(\Hil_c)^{\mathstrut}$}}
\put(7.5,2.5){\dashbox{2}(75,30)[lb]{ Quantum bus
 $(\Hil_d)_{\mathstrut}$}}
\end{picture}
\caption{Hybrid quantum control;
{\normalsize $\op\theta_k = e^{i \op\sigma_k \varphi}$}, where $\varphi$ is
 continuous parameter of control and {\normalsize $\op\sigma_k$} is Pauli matrices.
 Wavy (\wave{.5cm}) and straight (---) lines are continuous and
 finite quantum variables respectively ({\em cf} \cite{Vla02}).}
\label{fignet}
\end{center}
\end{figure}

One simple method of description of quantum computation with continuous
variables is to consider some functional spaces and space of linear
differential operators. Well known example are operators of coordinate
and momentum $\op p$, $\op q$.
Draft of universal quantum controller, based on design of hybrid quantum
processor \cite{Vla02} depicted on Fig.~\ref{fignet}.

Here controlled system is anyway finite-dimensional and only some subset
of controlling variables described by infinite-dimensional Hilbert space.
Interesting question is problem of universal control of continuous variables.
Such models were described yet only in semiclassical approach to quantum
computation and control. It was shown, that Hamiltonian with (bi)linear
combinations of coordinate and momentum are not enough \cite{LlBr98} for
universal quantum computation (control) and so some nonlinear (third-order)
\cite{LlBr98} or exponential \cite{VlaQI02} expressions may be used instead.

It is clear from previous consideration, that it is simpler to use some
analogue of universality in approximate sense for control of quantum
continuous variables --- it was discussed earlier, {\em dimension of Hilbert
space for universal control must coincide with cardinality (``number of
points'') in Hilbert space of controlled system} and so for controlled system
with continuous variables (\contr{\infty^\infty}\infty) such idea would
produce very different problems related with mathematical theory of
infinite cardinal numbers.

So, (precisely) universal control of $N$-dimensional system is possible
using continuous quantum variables (\contr\infty N), but quite
likely, that control of continuous quantum system (\contr\infty\infty)
may be universal only in approximate sense.
On the other hand, distinction between approximate and rigor universality
in last case has rather theoretical significance, because it is not
clear, how to find a difference between such \contr\infty\infty-controllers
during any {\em finite amount of time}. Anyway, both tasks discussed below
are difficult and out of scope of presented paper.

In addition, more accurate consideration of models of quantum computations
and control with continuous variables is not complete without
necessary attention to principles of quantum field theory.
This difficult area is still in state of development,
especially because correct description of quantum fields is possible only
using relativistic theory \cite{Vla96,PerTer02,TernoPhD}.


\end{document}